# Records of Auroral Candidates and Sunspots in *Rikkokushi*, Chronicles of Ancient Japan from Early 7th Century to 887

Hisashi Hayakawa (1–2)*, Kiyomi Iwahashi (3), Harufumi Tamazawa (4), Yusuke Ebihara (5–6), Akito Davis Kawamura (4), Hiroaki Isobe (6–7), Katsuko Namiki (8), Kazunari Shibata (4)

(1) Graduate School of Letters, Osaka University, Toyonaka, Japan
(2) Research Fellow of Japan Society for the Promotion of Science, Tokyo, Japan
(3) National Institute of Japanese Literature, Tachikawa, Japan
(4) Kwasan Observatory, Kyoto University, Kyoto, Japan
(5) Research Institute for Sustainable Humanosphere, Kyoto University, Uji, Japan
(6) Unit of Synergetic Studies for Space, Kyoto University, Kyoto, Japan
(7) Graduate School of Advanced Integrated Studies in Human Survivability, Kyoto University, Kyoto, Japan
(8) Kyorin University, Mitaka, Japan.

*E-mail: hayakawa@kwasan.kyoto-u.ac.jp

**Abstract**
In this article, we present the results of the surveys on sunspots and auroral candidates in *Rikkokushi*, Japanese Official Histories from the early 7th century to 887 to review the solar and auroral activities. In total, we found one sunspot record and 13 auroral candidates in *Rikkokushi*. We then examine the records of the sunspots and auroral candidates, compare the auroral candidates with the lunar phase to estimate the reliability of the auroral candidates, and compare the records of the sunspots and auroral candidates with the contemporary total solar irradiance reconstructed from radioisotope data. We also identify the locations of the observational sites to review possible equatorward expansion of auroral oval. These discussions suggest a major gap of auroral candidates from the late 7th to early 9th century, which includes the minimum number of candidates reconstructed from the radioisotope data, a similar tendency as the distributions of sunspot records in contemporary China, and a relatively high magnetic latitude with a higher potential for observing aurorae more frequently than at present.

**1. Introduction**
Information of long-term solar activity is important in the viewpoint of not only solar physics but also the relationship between solar activity and the terrestrial climate. Modern scientific data of solar activity are based on sunspot observations since the early 17th century (Hoyt & Schatten 1998; Owens 2013; Clette et al. 2014; Savalgaard & Schatten 2016). On the contrary, in order to study the solar activity during the pre-telescopic age, there are two types of proxies: the ratio of radioactive isotopes and the records of sunspots and aurorae in historical documents (Eddy 1980; Vaquero & Vázquez 2009). Radioactive isotopes such as Carbon-14 and Beryllium-10 are used to reconstruct the group sunspot number or total solar irradiance (TSI) during the pre-telescopic age (e.g., Solanki et al. 2004; Steinhilber et al. 2009). Conversely, historical records of sunspots and aurorae can be used as another proxy for solar activity during the pre-telescopic era (e.g., Vaquero & Vázquez 2009; Riley et al. 2015).

The largest solar event in the history of ground-based telescopic observations is thought to be the so-called Carrington event in 1859 (Carrington 1859). An unusually large magnetic change was recorded at low latitudes, which reached about −1600 nT. The disturbance storm index (Dst) was estimated to be −1760 nT (Tsurutani et al. 2003) and −625 nT (Siscoe et al 2006). Carrington (1859; 1863) reported a large sunspot coupled with a white flare in his study on 01 September 1859. This flare caused worldwide auroral observations even in low-latitude areas such as Hawaii, the Caribbean coasts, and Japan on the following day, 02 September 1859 (Kimball 1960; Tsurutani et al. 2003; Cliver & Svalgaard 2004; Green & Boardsen 2006; Cliver & Dietrich 2013; Lakhina & Tsurutani 2016; Hayakawa et al. 2016b). This case eloquently reveals that it is possible to trace back the extreme events and solar activity during the pre-telescopic age with records of aurorae and sunspots. Recent studies reveal that historical documents with datable descriptions allow us to trace back historical solar activity with literal auroral records up to 567 BCE (Stephenson et al. 2004; Hayakawa et al. 2016c), auroral drawings up to 771/772 (Hayakawa et al. 2017b), literal sunspot records up to 165 BCE (Yau & Stephenson 1988), and sunspot drawings up to 1128 (Stephenson & Willis 1999; Willis & Stephenson 2001). At the same time, several authors published catalogs of naked-eye sunspots and aurorae in historical documents from Japan (Kanda 1933; Matsushita 1956; Nakazawa et al. 2004), Korea (Lee et al. 2004), West Asia (Basurah 2006; Hayakawa et al. 2017b), Europe (Fritz 1873; Link 1962; Dell'Dall'Olmo 1979; Stothers 1979; Vaquero & Trigo 2005; Vaquero et al. 2010), Russia (Vyssotsky 1949), the Tropical Atlantic Ocean (Vázquez & Vaquero 2010), and China (Keimatsu 1970–1976; Yau & Stephenson 1988; Yau et al. 1995; Hayakawa et al. 2015, 2016a, 2017c; Kawamura et al. 2016; Tamazawa et al. 2017).

Recent studies suggest that even larger solar activity potentially occurs. Maehara et al. (2012) reported that flares with much larger energies (superflares) than the Carrington event occur on many solar-type stars, i.e., slowly rotating G-type stars. The total energy of superflares is estimated to be 10−1000 times as large as the Carrington event. Miyake et al. (2012, 2013) discovered an anomalous increase in Carbon-14 in tree rings, which indicates that some event caused a sharp increase in comic-ray flux during 774/775 and 993/994. These events attracted scientific interest, prompting scientists to propose several candidates for their origins such as extreme solar events (e.g., Usoskin et al. 2013; Cliver et al. 2014; Stephenson 2015; Mekhaldi et al. 2015; Hayakawa et al. 2016a, 2017a; Tamazawa et al. 2017).

In this article, we present the results of a survey of *Rikkokushi*, Japanese ancient Official Histories from the early 7$^{th}$ century to 887, to renew and recompile the Japanese auroral catalogs in the ancient times, a period contemporary to the *Suí* and *Táng* dynasties of China (Tamazawa et al. 2017). Japanese auroral catalogs during the pre-telescopic age have hitherto been compiled by some astronomers (Kanda 1933; 1934; 1935; Matsushita 1956; Nakazawa et al. 2004). Within these catalogs, Kanda (1933, hereafter K33) and Matsushita (1956, hereafter M56) compiled catalogs of ancient Japan covered by *Rikkokushi*. K33 listed two events (A1 and A2 in our catalog) and M56 listed four events (A1–A3 and one omitted in our catalog). However, we must consider two recent changes in the Japanese archaeoastronomical studies. First, comparison with Chinese astronomical records let us find not only "*sekki*/red vapor" (赤氣) but also other terms such as vapor (氣), cloud (雲), and light (光), which could indicate auroral candidates (Keimatsu 1970–76; Hayakawa et al. 2015, 2016b). Hayakawa et al. (2016a) examined whether astronomical records described as "white/unusual rainbows" include auroral candidates, which was also supported by Carrasco et al. (2017). Secondly, these records of sunspots and auroral candidates have not been compared with contemporary solar activity reconstructed from scientific datasets. Therefore, we compared the survey results from *Rikkokushi* with solar activity data such as the TSI or historical records in other regions. These results are of much importance to consider contemporary solar activity such as the other grand minimum candidate from 640 to 710 reported by Eddy (1977) and Usoskin et al. (2007).

## 2. Method
### 2.1. Astronomical Observations in Japan
In order to compile a comprehensive catalog of ancient aurorae in Japan, we consult *Rikkokushi*, a series of Japanese official histories[1] consisting of *Nihonshoki* (NS, 日本書紀)[2], *Shokunihongi* (SNG, 続日本紀)[3], *Nihonkoki* (NHK, 日本後紀)[4], *Shokunihonkouki* (SNK, 続日本後紀)[5], *Nihon*

---

[1] After NSJ, no official histories were compiled in Japan. It has been difficult to compile the Japanese auroral catalogs after 888. We are going to discuss this matter in another paper.
[2] Spans from the mythological era to 697, compiled in 720.
[3] Spans from 697 to 791, compiled in 797.

*Montoku Tennou Jitsuroku* (NMTJ, 日本文徳天皇実録)[6], and *Nihon Sandai Jitsuroku* (NSJ, 日本三代実録)[7]. We examined the wood print versions of the 17th century currently stored at the National Archives of Japan. The references of these source documents are presented in Appendix 1, and the references of other historical documents for additional discussion are presented in Appendix 2.

These official histories were compiled under strong influence from *Táng* (唐), the Chinese contemporary dynasty (Sakamoto 1970). In this period, *Táng* was a dominant power in the whole of East Asia and had diplomatic leadership in the Eastern Asian international relationship. East Asian countries tried hard to import and imitate Chinese systems and cultures such as how to rule their country, build their capital city, write, and record their official history (正史). The Japanese official histories were written in Chinese and their formats were imported from those of the Chinese imperial chronicles (本紀) in the official histories (正史) (Sakamoto 1970; Tamazawa et al. 2017).

*Onmyoryo* (陰陽寮) under *Nakatsukasa-sho* (中務省), the ministry of inner affairs, engaged in astronomical observations and reported their results. When finished, they categorized these ominous records and sent them to *Nakatsukasa-sho* to enter into the national histories (*Ryonoshuge*, p.334). It was at the new-year celebration of 675 (on 01 February 675) (NS, v.29, f.5a) that *Onmyoryo* appeared in the Japanese historical documents for the first time. It is stated that the students at *Daigakuryo* (大学寮), staffs at the department of the medical office, and *Onmyoryo*, girls of *Sawe* (舍衞女), girls of *Tara* (堕羅女), the defected king of Baekje (百済/Kudara), servants of Silla (新羅/Shiragi)[8] came to offer medicine and treasures. This was the beginning of "*Migusuri* (御薬)," an annual celebration where the Japanese emperor and his subordinates wished for longer lives. Before that, we have a report that a monk from Baekja brought "documents of almanac, astronomy, and geography" into the Japanese court and the Japanese government selected some students to master these arts in 602 (NS, v.22, ff.5a–b). Tanikawa & Soma (2008) and Soma & Tanikawa (2011) analyzed the early astronomical records in *Rikkokushu* and concluded that the astronomical records up to "red vapor (赤氣)" in 620 were at least based on contemporary astronomical observations.

**2.2. Survey Method**
The ancient Japanese astronomers recorded auroral candidates and sunspots in terms such as "red vapor" (see Kanda 1933) or "black spots". These terms were literally derived from the contemporary Chinese official histories. Thus, we regard the other terms in the Chinese official histories such as "vapor (氣)," "cloud (雲)," and "light (光)" with color (Keimatsu 1970–76; Yau et al. 1995; Willis et al. 1999; Hayakawa et al. 2015, 2016b) as auroral candidates, and "black spot (黒點)" as sunspot candidates in the Japanese official histories as well to make a machine retrieval in the entire text of *Rikkokushi*. We converted the date shown in Japanese luni-solar calendar to those in Julian calendar based on Uchida's conversion tables (Uchida 1992, 1993). Note that the boundary of contemporary date was placed around 23:00 (Soma et al. 2004)

**3. Results and Discussion**
**3.1. Overall Result**
In total, we found one sunspot record and 13 auroral candidates (vapor/cloud/light with color) with events seen during daytime excluded. The original texts and translations are presented in Appendices 3 and 4, and their summarized catalogs of auroral candidates are presented in Table 1. Among the 13 auroral candidates, six are described as white, six are described as red, and one is described as bluish[9]. Moreover, four include vapors, eight include clouds, and one includes light. The catalog is also available via our website: http://www.kwasan.kyoto-u.ac.jp/~palaeo/.

---

[4] Spans from 792 to 833, compiled in 840.

[5] Spans from 833 to 850, compiled in 869

[6] Spans from 850 to 858, compiled in 879.

[7] Spans from 858 to 887, compiled in 907

[8] Baekja (百済) and Silla (新羅) are contemporary Korean dynasties.

### 3.2. Sunspot Records
Within the datasets of *Rikkokushi*, we found one sunspot record from 02 December 851 (**Appendix 3, S1**).

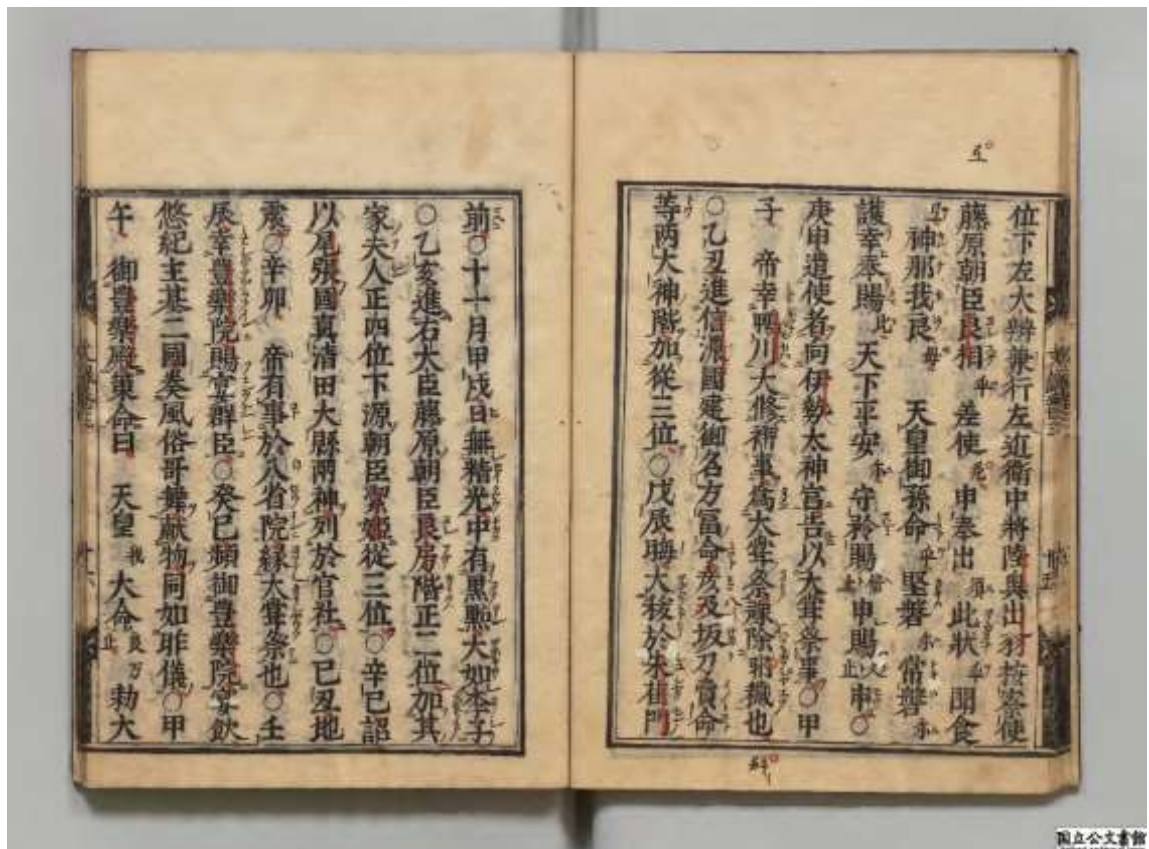

Figure 1: The earliest sunspot in record from 851 within NMTJ (v.3, f.16a).

Based on its description, we can easily find similarities with sunspot records in the Chinese official histories. Its size was compared to a plum fruit, as seen in the Chinese and Korean official histories (e.g., Lee et al. 2004; Hayakawa et al. 2015; Tamazawa et al. 2017), although it is recorded in a different character (in Japanese 黒點; in Chinese 黒子). Considering that this sunspot was reported to be observed when "the sun was not bright," we may estimate the presence of Asian dust (黄砂/黄沙) to allow astronomers to observe this sunspot.

### 3.3. Auroral Candidates and their descriptions
We found 13 auroral candidates, whereas K33 and M56 only reported two events (A1 and A2) and four events (A1–A3 and one excluded in our article), respectively, as they seem to have only surveyed "vapors (氣)" with color. Note that we do not include the event on 21 April 843 reported by M56, which was reported to have been seen around 08:00 (食時)[10] (SNK, v.13, f.11b), clearly during the daytime, and hence out of our criteria. K33 includes one red vapor (A1, see Figure 2) and one white vapor (A2) in NS. M56 includes two red vapors in SNK. We adopt one of them as A3, while we omit the other of 21 April 843. We found one more "vapor" around the moon (A3). Although that might be related to the lunar halo, we must note that the record itself also addresses the lunar halo (月暈) and their motions independently in the same sentence, and Japanese astronomers followed the tradition of Chinese astronomy, which possessed so deep knowledge about halos that they had 26 technical terms for them (Ho & Needham 1959).

---

[9] As for white/unusual rainbow in *Rikkokushi*, the detail is discussed in Hayakawa et al. (2016a).

[10] While K56 adds a note that it was seen around 16:00, it should be addressed not as *shokuji* (食時) but as *hoji* (晡時).

According to Chinese terminology, we have records with "lights (光)" and "clouds (雲)" with color (A4–13). We have three red clouds (赤雲), one red light (赤光), one bluish cloud (青雲) and five white clouds (白雲). These terminologies are used in the official histories of the *Táng* dynasty (Tamazawa et al. 2017) and for simultaneous auroral observations caused by great magnetic storms such as those in 1363 or the Carrington event in 1859 (Willis & Stephenson 1999; Hayakawa et al. 2016a, 2016b).

Motions of these phenomena are provided by eight points of the compass. Sometimes, descriptions for these phenomena also include information about length and appearance. Their length is given in *shaku* (尺) or *jou* (丈). One *jou* equals ten *shaku* and one *shaku* in ancient Japan equals 30.3 cm (*Ryonogige*: p.333). Their appearance is sometimes compared with "lightning," "fire," "blanket" or "*kanjo no hata*" (Figure 3) to describe their shapes or brightness.

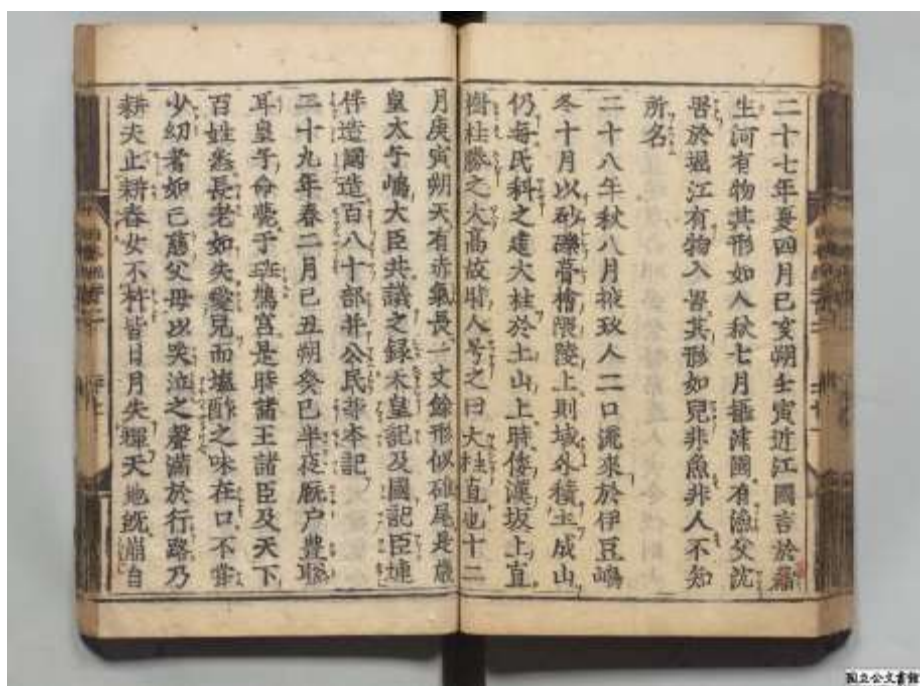

Figure 2: The earliest record of auroral candidate in 620 in NS (A1)

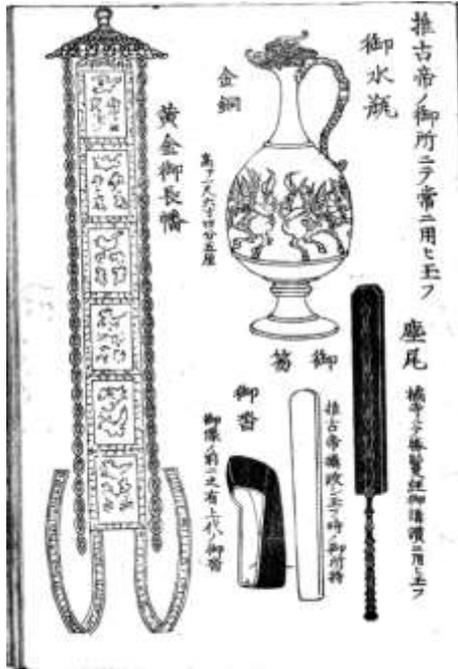

Figure 3: *Kanjo no hata* is a type of long suspended cloth, a Buddhist altar fitting. In ancient times, it was an open work on a bronze tablet. They were in use in the Imperial palace even since Empress Suiko (r.593–628) and one of them has been preserved in the Horyu Temple, as described in *Horyu-ji Houmotsu Zue* (f.4a).

The color of the auroral candidates in *Rikkokushi* is described as red, white, and blue, as explained above. Normally, the red color of aurorae comes from the spectral line of atomic oxygen at 630.0 nm (red line), excited primarily at high altitudes (> 150 km), and the green color is represented by the spectral line of atomic oxygen at 557.7 nm (green line), primarily at middle altitudes (100–200 km). The visible, blue color probably comes from the spectral lines of molecular nitrogen ions, for example, at 391.4 and 427.8 nm, which can sometimes be seen above the green aurora (Chamberlain 1961). At the lower border of an auroral curtain (90–100 km), the purple color sometimes dominates, which is caused by the combination of the red and blue spectral lines of molecular nitrogen ions. The auroral candidates in *Rikkokushi* include a blue one. The blue color may also come from the spectral line of hydrogen at 486.1 nm, the so-called proton aurora. Proton aurorae are frequently seen equatorward of the main auroral oval and are caused by the precipitation of protons. However, their brightness is usually too low to be visible by the naked eye. In addition to the spectral line of hydrogen at 486.1 nm, proton precipitation may result in the intensification of the green line (Sakaguchi et al. 2007). Some of the red aurorae described in *Rikkokushi* may correspond to stable auroral red arcs (SAR arcs). SAR arcs are excited by thermal conduction from the hot inner magnetosphere to cold ionosphere whose color is exclusively red. In *Rikkokushi*, their forms are sometimes compared to something long: pheasant's tail (A1), kanjo no hata (A2), flag-cloud (A7), and snake (Ra1). Likewise, some of their directions are described as "crossing the heavens (竟天)" (A6) or crossing from the east to west (A5, A6, A11, A13). Some records include descriptions of length (A1, A3, A5, A8, A12). A10 and A13 appear to move westward. Their shape may indicate the curtain-like (arc-like) shape of aurorae rather than the diffusive shape of aurorae. During large magnetic storms, SAR arcs can be very bright (~13 kilo Rayleighs), but their structure appears fairly broad (Baumgardner et al. 2007). Thus, we suppose that some of them may be different from SAR arcs. Usually, curtain-like aurorae are associated with upward field-aligned currents (FACs) that usually flow into and away from the polar ionosphere at magnetic latitudes (MLATs) of 65–70°. For the large magnetic storm of 20–21 November 2003, strong upward FACs were observed at approximately 45° in MLAT (Ebihara et al 2005). We cannot exclude the possibility that the upward FACs might have been able to flow at about 35° in MLAT. Further studies are needed to verify this possibility. The white aurorae may be explained as follows. In such a case that the auroral lights are too faint to be detected by the cone cells of the human eye responsible for color vision, the rod cells of the human eye used in peripheral vision dominate because of the so-called Purkinje effect (Purkinje 1825) and, hence, green aurorae can be observed as being whitish (Tamazawa et al. 2017). It is also possible that some of the strong green aurorae are described as bluish because of the color recognition in ancient East Asia (e.g., Pankenier 2013), although it is also possible to explain the bluish aurorae by the excited nitrogen or hydrogen (Tinsley et al. 1984).

Some records compare their intensity with lightning (A7, A8) or torchlights (A3), which cannot be explained very well by atmospheric optics or halos. Their observational time is not always described. Four auroral candidates are recorded to be observed during the night (A3, A4, A8, A12), two auroral candidates at sunrise (A4, A11), and one auroral candidate at sunset (A11). It is known that intense aurorae under extreme magnetic storms can be observed even in twilight (Loomis 1860; Willis & Stephenson 1999; Hayakawa et al. 2016b).

**3.4. Aurora and Moon Phase**
All the auroral candidates shown here do not necessarily imply that they are definitely aurorae, as we must expect some contamination from atmospheric optics or paraselenes (Hayakawa et al. 2015, 2016a; Kawamura et al. 2016; Usoskin et al. 2017). In the night sky, the largest source of light contamination is usually that of the moon. Hence, we computed the normalized moon phase of every event according to the method of Kawamura et al. (2016) developed from the algorithm of "moonphase.pro" in the IDL Library of NASA based on Meeus (1998).

Figure 4 shows distributions of the auroral candidates against the normalized lunar phase. We could not find any meaningful interpretation of this histogram. Three of them are found around the new moon and are hence considered to receive little influence from the moonlight: A1 and A13 fall into the lunar phase < 0.1 while A3 falls into the lunar phase > 0.9. However, we must note that bright aurorae can be observed even with the presence of the full moon and, hence, we cannot simply exclude the auroral candidates that appear around the full moon (see Loomis 1860; Bernard 1910; Tamazawa et al. 2017).

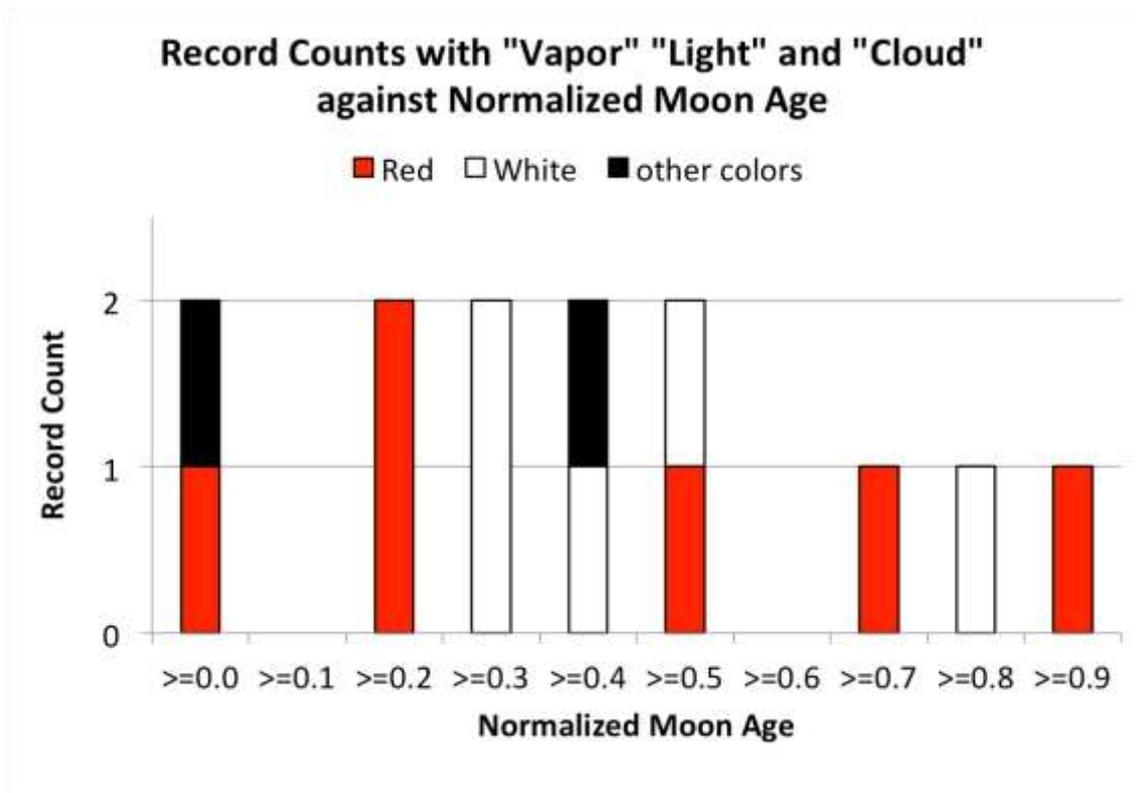

Figure 4: Record counts with "Vapor," "Light," and "Cloud" against normalized moon phase

### 3.5. Comparison with Radioisotope Data

It is intriguing to compare the records of auroral candidates and sunspots in *Rikkokushi* with the index of long-term solar activity. Here, we compare the records of auroral candidates and sunspots in *Rikkokushi* with the TSI reconstructed from carbon-14 and beryllium-10 by Steinhilber et al. (2009) and sunspot records from the contemporary Chinese official histories, *Jiùtángshū* and *Xīntángshū*, plotted by Tamazawa et al. (2017). We plotted the records of auroral candidates and sunspots in Rikkokushi against TSI in Figure 5a, sunspot records of the contemporary Chinese official histories against TSI in Figure 5b, and both against TSI in comparison with the longer-term TSI transition from 1–2000 in Figure 5c.

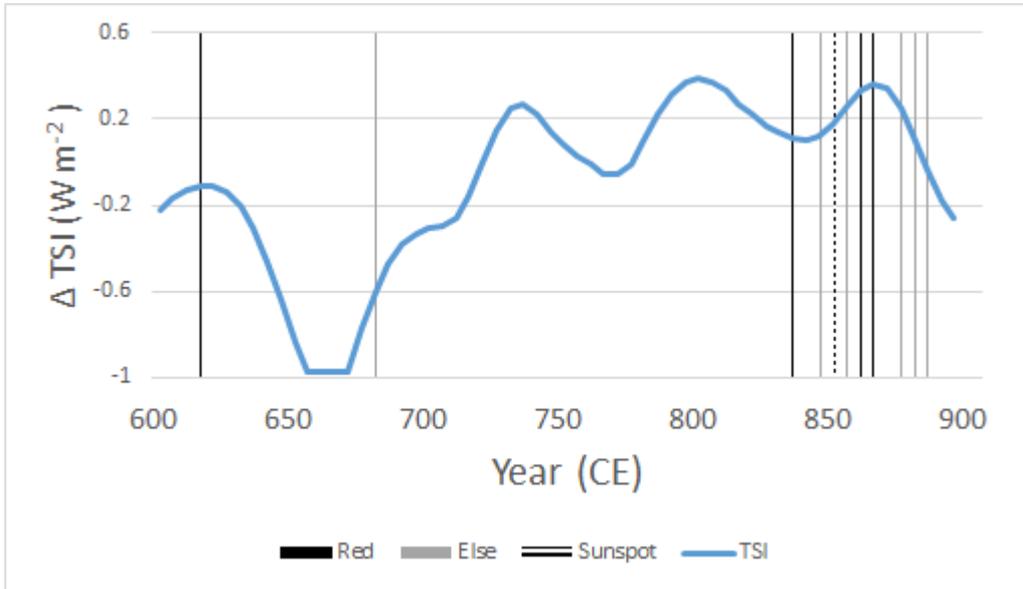

Figure 5a: Records of sunspots and auroral candidates in *Rikkokushi* in comparison with TSI by Steinhilber et al. (2009)

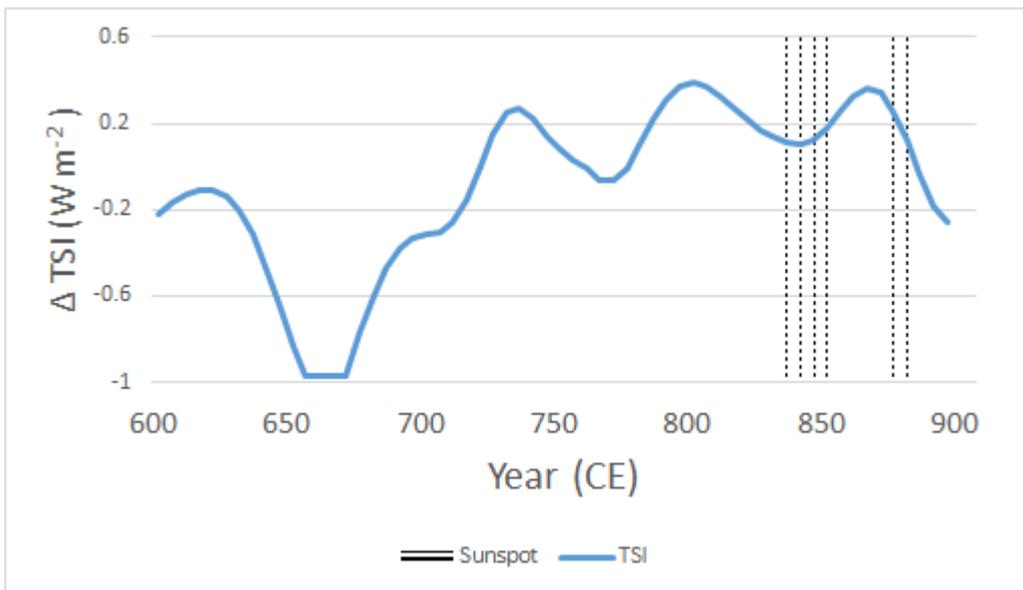

Figure 5b: Records of sunspots in contemporary Chinese official histories, *Jiùtángshū* and

*Xīntángshū* (Tamazawa et al. 2017), in comparison with TSI by Steinhilber et al. (2009)

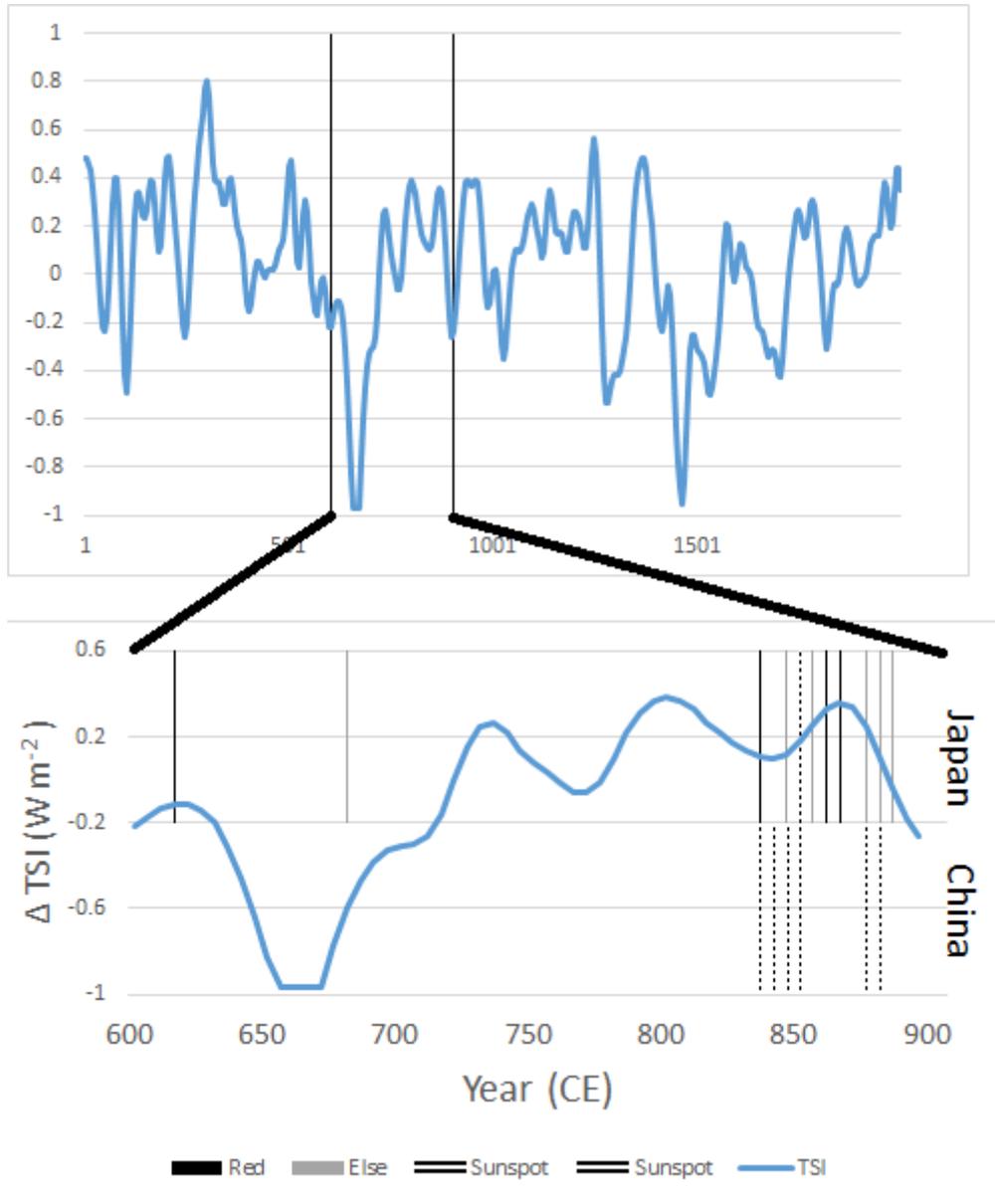

Figure 5c: Records of sunspots and auroral candidates in *Rikkokushi* in comparison with the TSI from 1 to 2000 by Steinhilber et al. (2009), and records of sunspots in contemporary Chinese official histories.

In Figure 5a, we can find a major gap between 620 (A1) and 839 (A3), except for that in 681 (A2), which includes the candidate of a grand minimum from 640 to 710, as reported by Eddy (1977) and Usoskin (2007). We may partially explain the great absence from 710 to 839 by the potential lack of intensive observations in the Nara period, as it is known that most records of solar eclipse are predictions in this period (Takesako 2011). On the other hand, it is notable that most of the auroral candidates occur after 839 with the sole sunspot record in 851.

Figures 5b and 5c suggest that this tendency of clustered auroral records after the mid-9[th] century is also found in the contemporary Chinese sunspot records between 835 and 875, as reported by Tamazawa et al. (2017). Auroral candidates A3–A4 are in the declining phase in the mid-840s. Auroral candidates A6–A10 are in the developing phase at the peak in the 870s. Auroral candidates

A11–A14 are in the declining phase in the 890s with the former two near the peak in the 870s. Especially, A3 corresponds with the German auroral record in 839 (Link 1962) and A10 with the Chinese sunspot record in 865 (Tamazawa et al. 2017), suggesting strong solar activities in these years.

A comparison with the longer-term TSI (1–2000) reveals that this period covered by *Rikkokushi* (c.a. 600–887) coincides with the start of the declining phase from the early 6th century. Around 620 (A1), the TSI recovers by a small amount before sharply decreasing into the minimum candidate during 640–710 (Eddy 1977; Usoskin 2007). The TSI suggests that the solar activity during the candidate of the grand minimum was even lower than that of the Maunder grand minimum and possibly the lowest in the last 2000 years. Conversely, the TSI during the 9th century suggests that the contemporary solar activity was mostly in the active phase, possibly as much as that of the Medieval Maximum during 1100–1250.

### 3.6. Location of Observational Sites

In order to evaluate auroral activities, we need to know where these observations were carried out (e.g., Hayakawa et al. 2016b). Only M56 identifies the previous Japanese observational sites: A1 and A2 at Nara, and A3 and the event of 21 April 843 at Kyoto. Here, we examine the locations of the contemporary observational sites based on where the observatories were placed and correct some of M56's interpretation.

It is known that *Onmyoryo* engaged in astronomical observations with calendar compilation since 701 (*Ryonogige*, p.36). *Onmyoryo*'s duty consisted of seeing the yin and yang (陰陽), calendars, astronomy, and time calculation. It was quite unique how *Onmyoryo* existed in Japan. Although Japanese *Ritsuryo* (律令) i.e., contemporary laws, were under overwhelming influence from those of the *Táng* (唐) dynasty in China, *Onmyoryo* and *Jingikan*, i.e., Japanese professional department of Shintoism, had unique characteristics. In *Táng*, those duties were divided into *Tàibǔshǔ* (太卜署) for astro-omenology and *Tàishǐjú* (太史局) for compiling calendars, astronomical observation, and calculating time, while in Japan, those duties were integrated into one department with a focus on fortune-telling. This is why superstition later had much influence on astronomy and calendars in Japan (Murayama 1991).

Their *senseidai* (占星台), i.e., astronomical observatories, were set up on 05 February 675 for the first time, four days after the first appearance of *Onmyoryo* in NS (v.29, f.5a). *Senseidai* corresponds to *língtái* (靈臺) or *sītiāntái* (司天臺) in Chinese traditional astronomy (Arakawa 2001). The fact that "Japanese observatory" has the name of "*sensei* (占星)" suggests that astro-omenology had a considerable amount of influence on contemporary Japanese astronomy as well. Because of the lack of detailed records, it is difficult to reconstruct what type of instrument or equipment were in use in ancient Japanese astronomy.

The naked-eye observatories are not necessarily in the mountaneous areas. So, the ancient and medieval observatories were located near the capital[11] (e.g., Keimatsu 1976; Pankenier 2003; Hayakawa et al. 2015; Tamazawa et al. 2017). Japan followed this tradition and hence we can approximately estimate where the observational site was according to where the contemporary capital was (see Table 2). Until 674, the Japanese capital was placed at Asuka Kyo and the remains of the buildings with water pipes and the tanks of ancient water clocks excavated at Site Mizuochi (水落遺跡: N34°29′, E135°49′) in Asuka Kyo are estimated to be from the contemporary astronomical observatory (Kinoshita 1987; Arakawa 2001). At Heijo Kyo (平城京, current Nara), during 710−740 and 745−784, *Onmyoryo* was placed on the east side of the capital, which is currently known as In'yo-cho (陰陽町) after *Onmyoryo*'s activity. At Heian Kyo (平安京, current Kyoto), after 794, *Onmyoryo* was placed on the southern side of the Imperial Palace. On 23 March 1127, a great fire destroyed everything but the astrological tables and water clocks (*Chuyuki*: p.374). No more records on the observatories or their reconstruction since that time are available down to the Kyoto Umenokouji Observatory (京都梅小路天文台), which was constructed by Abe no Yasukuni (安倍泰邦) from the Tsuchimikado (土御門) Family in 1751 (*Jujikai*, v.15, ff.3a-b; Hashimoto 1976; Watanabe 1987). This may suggest that professional observatories may not have been reconstructed after that fire. Moreover, the revision of calendars had long been abandoned since the Senmyo Calendar (宣明暦) was adopted in 862 and the *Jokyo* Calendar was adopted in 1685 (Yuasa 2009).

---

[11] The earliest surviving observatory in East Asia is *Cheomseongde* (瞻星台) in *Gyeongju* (慶州). It exists at the southern end of the ancient capital of *Gyeongju* in the era of *Silla* (新羅).

Therefore, we identify the observational sites of A1 and A2 as Asuka and A3 to A13 as Kyoto. This implies that we need to correct M56's interpretation for observational sites A1 and A2.

### 3.7. Magnetic Latitude in Japan

Identifying the observational sites allows us to compute the magnetic latitude with the aid of a proper geomagnetic field model. Therefore, we computed the magnetic latitudes of Asuka Kyo and Heian Kyo (Kyoto), which are the southernmost and northernmost capitals of contemporary Japan, respectively, and the results are presented in Figure 6. Here, we defined the magnetic latitude in accordance with the angular distance from the dipole axis and used the global geomagnetic field model CALS3k.4b by Korte & Constable (2011) to calculate the location of the dipole axis.

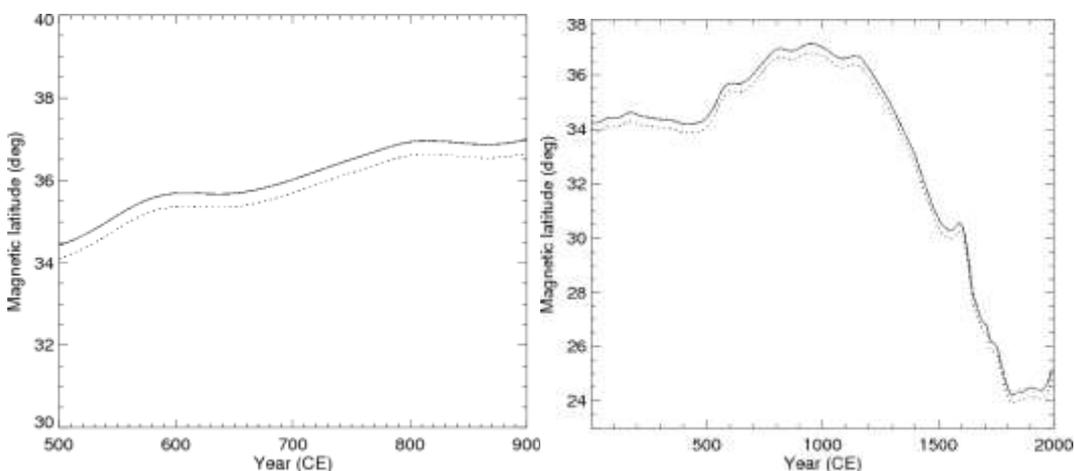

Figure 6: Magnetic latitudes of contemporary observational sites of Heian Kyo (solid line) and Asuka Kyo (dotted line) during the periods of 600–887 (Figure 6a) and 1–2000 (Figure 6b).

Figure 6a shows that the magnetic latitudes of the Japanese capitals, Heian Kyo and Asuka Kyo, were between about 35 and 37° in MLAT during the period between the early 7$^{th}$ century and 887. The magnetic latitudes were about 10° higher than those of present-day and were almost the highest throughout the recent two millennia. If the aurora was seen at 10° from the horizon and the height of the red aurora was at an altitude of 400 km, the equatorward boundary of the auroral oval would extend to about 47 and 49° in MLAT. Considering the observational fact that the equatorward boundary of the auroral oval correlates with the scale of the magnetic storm (Yokoyama et al. 1998), we estimated the Dst index to be –282 to –211 nT. This class of magnetic storm frequently occurs near solar maxima. It should be emphasized that this is the lower limit because the aurora was likely observed at elevation angles larger than 10°, and the magnetic activities captured in these official histories may have been more extreme. This estimation is consistent with the AUREST (AURora ESTimation) model, predicting that Heian Kyo and Asuka Kyo were within the border where aurorae were visible on the horizon at Kp = 9 (Korte & Stolze, 2014).

Figure 6b also indicates that the magnetic latitudes of these cities started to fall because of the shift in the geomagnetic dipole toward America after the mid-12$^{th}$ century. The magnetic latitudes of these cities have been 24–25° since the 19$^{th}$ century, so aurorae are exclusively visible in Japan during extreme magnetic storms. Aurorae during the Carrington magnetic storm of 1859 were seen in Inami or Shingu at 23° in MLAT (Hayakawa et al. 2016b).

## 4. Conclusions

We have surveyed the auroral candidates in *Rikkokushi*, six Japanese Official Histories up to 887. We found one sunspot in 851 and 13 auroral candidates during this period, with information of their motion, direction, shape, and color. The catalog is presented in this paper. By computing the contemporary magnetic latitudes of the observational sites, we found that they were auroral observations from the observational sites at 35−37° in MLAT. The absence of auroral records support the existence of a weak grand minimum between 620 and 720 suggested by radioisotope proxies, and the active phase after 835 was also found by the isotope proxies, as well as the sunspot and auroral records in contemporary China. This is one in a series of our auroral surveys on historical documents and the first step in Japanese historical documents to provide online data for use by the scientific community. We welcome and encourage the use of our data as well as the contributions that provide more data from diverse historical sources.

## Acknowledgements

We acknowledge the support of Kyoto University's Supporting Program for the Interaction-based Initiative Team Studies "Integrated study on human in space" (PI: H. Isobe), the Interdisciplinary Research Idea contest 2014 by the Center for the Promotion of Interdisciplinary Education and Research, the "UCHUGAKU" project of the Unit of Synergetic Studies for Space, the Exploratory and Mission Research Projects of the Research Institute for Sustainable Humanosphere (PI: H. Isobe) and SPIRITS 2017 (PI: Y. Kano) of Kyoto University, and the Center for the Promotion of Integrated Sciences (CPIS) of SOKENDAI. This work was also encouraged by a Grant-in-Aid from the Ministry of Education, Culture, Sports, Science and Technology of Japan, Grant Number JP15H05816 (PI: S. Yoden), JP15H03732 (PI: Y. Ebihara), JP16H03955 (PI: K. Shibata), and JP15H05815 (PI: Y. Miyoshi), and Grant-in-Aid for JSPS Research Fellow JP17J06954 (PI: H. Hayakawa). H. H. thanks the National Archive of Japan for providing the wood print edition of *Rikkokushi*, Dr. K. Tanikawa for his helpful comments on our manuscript, Ms. C. Kuroyanagi for helps to access some relevant articles, Dr. R. Kataoka and Dr. H. Miyahara for the helpful advice for interpretations on the records of auroral candidates, and Dr. T. Terashima for providing insights and advice on technical terms in *Rikkokushi*. Y. E. thanks Dr. M. Korte for providing us with the geomagnetic field model and related computer programs.

**Appendix 1: References of *Rikkokushi***
NS: *Nihonshoki* (日本書紀), MS 特 127-0002, National Archives of Japan.
SNG: *Shokunihongi* (続日本紀), MS 137-0106, National Archives of Japan.
NHK: *Nihonkouki* (日本後紀), MS 270-0065, National Archives of Japan.
SNK: *Shokunihonkouki* (続日本後紀), MS 137-0132, National Archives of Japan.
NMTJ: *Nihon Montoku Tennou Jitsuroku* (日本文徳天皇実録), MS 137-0149, National Archives of Japan.
NSJ: *Nihon Sandai Jitsuroku* (日本三代実録), MS 特 025-0003, National Archives of Japan.

**Appendix 2: Bibliographic information of historical documents on the contemporary background of pre-telescopic astronomical observations.**
*Chuyuki*: Fujiwara-no-Munetada, *Chuyuki* (中右記) VII, Kyoto, Rinsen Shoten, 1965.
*Horyu-Ji Houmotsu Zue*: Horyu-ji, *Horyu-ji Houmotsu Zue* (法隆寺宝物図絵), MS Ya 8-34, National Insistute of Japanese Literature.
*Jujikai*: Nishimura-no-Tosato, *Jujikai* (授時解), MS 145, Library of National Astronomical Observatory of Japan.
*Ryonogige*: Kiyohara-no-Natsuno, Ryonogige (令義解, K. Kuroita, ed.), Tokyo, Yoshikawa Koubunkan, 1966.
*Ryonoshuge*: Koremune-no-Naomoto, *Ryonoshuge* (令集解, K. Kuroita, ed.), Tokyo, Yoshikawa Koubunkan, 1966.

**Appendix 3: Original texts and translations of sunspot records in *Rikkokushi***
Here, we present the original texts and translations of the sunspot records in *Rikkokushi*. Their ID, date, reference, original text, and translation are shown.

S1: 02 December 851, NMTJ, v.3, f.16a (see, Figure 1)
Original Text: 仁寿元年…十一月甲戌。日無精光。中有黒點。大如李子。
Translation: On 02 December 851, the sun was not bright and had a black spot as large as a plum fruit on its surface.

**Appendix 4: Original texts and translations of auroral records in *Rikkokushi***
Here, we present the original texts and translations of the auroral records in *Rikkokushi*. Their ID, date, reference, original text, and translation are shown.

A1: 30 December 620, NS, v.22, f.21b-22a (Figure 2)
Original Text: 推古天皇二八年…十二月庚寅朔。天有赤気。長一丈餘。形似碓（雉）尾。
Translation: On 30 December 620, a red vapor appeared in the heavens, and it was as long as 1 *shaku* and shaped like a pheasant's tail.

A2: 18 September 682, NS v.29, f.31a
Original Text: 天武天皇十一年[12]八月…壬申。有物。形如灌頂幡。而火色。浮空流北。毎國皆見。或曰。入越海。是日。白氣起於東山。其大四圍。
Translation: On 18 September 682, there was something like a *kanjo no hata*, whose color was like that of fire. It floated in the sky toward the north and observed every land of Japan and the Sea of *Koshi* (the Sea of Japan). Otherwise, it was said that in the day, a white vapor appeared on the eastern mountains, as large as the surrounding four sides.

A3: 08 October 839, SNK, v.8, f.14b
Original Text: 承和六年六月丁丑…是夜。有赤氣。方世丈。從坤方來。至紫宸殿之上。去地

---

[12] Note that the "reign of Emperor Tenmu (天武天皇)" in NS includes the period of Otomo-no-Oji who is sometimes known as "Emperor Koubun (弘文天皇)".

廿許丈。光如炬火。須臾而滅。

Translation: On 10 August 839, during the night, a red vapor appeared and was as long as and as wide as 30 *jou*[13]. It came from the southwest above *Shishinden*. It was 20 jou above the ground. Its light was like that of a torch. It disappeared after a while.

A4: 27 July 847, SNK, v.17, ff.11a-11b
Original Text: 承和十四年六月乙巳。…此夜。月暈之外有白氣繞之。
Translation: On 27 July 847, during the night, a white vapor emerged out of the lunar halo and surrounded it.

A5: 05 November 857, NMTJ, v.9, ff.27b-28a
Original Text: 天安元年十月己卯…是日。有白雲。廣四尺許。東西竟天。
Translation: On 05 November 857, during the day, a white cloud appeared and was as wide as about 4 *shaku*[14], extending across the heavens from the east to west.

A6: 24 July 858, NMTJ, v.10, f.18b
Original Text: 天安二年六月…庚子。早旦有白雲。自艮亘坤。時人謂之旗雲。
Translation: On 24 July 858, early in the morning, a white cloud extended from the northwest to southwest. Contemporary people called it a flag cloud.

A7: 13 November 859, NSJ, v.3, f.12a
Original Text: 貞觀元年…十五日丁酉。天東南有異雲。中有赤色。如電光激。
Translation: On 13 November 859, in the heavens toward the southeast, an unusual cloud appeared, as well as a red color as intense as lightning.

A8: 09 November 864, NSJ, v.9, f.10a
Original Text: 貞觀六年十月…七日庚申。夜。北山有光。如電。又朱雀門前見赤光。長五尺許。
Translation: On 09 November 864, during the night, a light appeared on the northern mountains and was as intense as lightning. Red light was also observed in front of Suzaku Gate and was as long as 5 *shaku*.

A9: 17 July 865, NSJ, v.11, f.2b
Original Text: 貞觀七年六月…廿一日庚午。遲明。月色正黃。有赤雲覆之。
Translation: On 17 July 865, around dawn, the color of the moon was purely yellow. There was a red cloud covering it.

A10: 28 August 876, NSJ, v.29, ff.7a-7b
Original Text: 貞觀十八年八月…六日庚戌。日入之時。赤雲八條起自東方。直指西方。廣殆及竟天。瑞祥志曰，天氣峙時。山川出雲。占云，赤氣如大道一條。若至三四五條者大赦。人民安樂。
Translation: On 28 August 876, at dusk, eight bands of red clouds appeared from the east and headed directly westward. It was so wide that it extended across the heavens. According to the treatises of omens, when vapor rises in the heavens, clouds appear from the mountains and rivers. According to omenologies, red vapor should appear as one band like a large road. In the case of the appearance of three, four, or five bands of them, grant a great amnesty and comfort to the people.

---

[13] While the woodprint edition of SNK gives "30 *jou*" here, we have some variants giving "40 *jou* (卅丈/四十丈)" such as manuscripts (MS ri05_02450: v.8, f.13a; MS ri05_08539: v.8, f.12a) in the Waseda University Library.

[14] "4 *shaku*" here might be an error of "4 *jou*," as "4 *shaku*" seems too short to be seen "extending across the heavens from the east to west." A woodprint in the National Diet Library (MS 839-5: v.9, ff.27b-28a) actually corrects "4 *shaku*" to "4 *jou* (四丈)".

A11: 16 October 876, NSJ, v.29, f.10a
Original Text: 貞觀十八年九月…廿五日己亥。天南有白雲。亙東西。
Translation: On 16 October 876, a white cloud appeared in the southern heavens, extending across from the east to west.

A12: 20 April 883, NSJ, v.43, f.7b
Original Text: 元慶七年三月…十日丙子。昏時月暈行犯大微西蕃上將星。亥時。白雲氣自北方來入暈中。其數五片。廣一尺許。長一丈。四片乃滅。一片貫月。良久消却。
Translation: On 20 April 883, at twilight, a lunar halo came into the Supreme Palace Enclosure, *Seiban* (Leo σ, ι, θ, δ) and *Joshosei* (Leo δ). Around 22:00, a white vapor emerged from the north and entered into the halo. It consisted of five pieces and was as wide as 1 *shaku* and as long as 1 *jou*. Four of them soon disappeared and one penetrated the moon and disappeared after a while.

A13: 14 August 885, NSJ, v.48, f.1b (Figure 6m)
Original Text: 仁和元年七月…卅日壬子。天有青雲。自東北竟西南。
Translation: On 14 August 885, a blue cloud appeared in the heavens, extending from the northeast to southwest.

**Table 1: Auroral candidates in *Rikkokushi***

| ID | Year | Month | Date | Color | Description | Direction | Place | Notes | Reference |
|---|---|---|---|---|---|---|---|---|---|
| A1 | 620 | 12 | 30 | R | V | | Asuka | like pheasant's tail | NS, v.22, ff.21b-22a |
| A2 | 682 | 9 | 18 | W, fire | V | | Asuka, all over Japan | like *kanjo no hata* | NS, v.29, f.31a |
| A3 | 839 | 8 | 10 | R | V | sw | Kyoto | like torchlight | SNK, v.8, f.14b |
| A4 | 847 | 7 | 27 | W | V | | Kyoto | around moon | SNK, v.17, ff.11a-11b |
| A5 | 857 | 11 | 5 | W | C | e-w | Kyoto | | NMTJ, v.9, ff.27b-28a |
| A6 | 858 | 7 | 24 | W | C | wn-en | Kyoto | flag-cloud | NMTJ, v.10, f.18b |
| A7 | 859 | 11 | 13 | R | C | | Kyoto | like lightning | NSJ, v.3, f.12a |
| A8 | 864 | 11 | 9 | R | L | es | Kyoto | like thunder | NSJ, v.9, f.10a |
| A9 | 865 | 7 | 17 | R | C | | Kyoto | near moon | NSJ, v.11, f. 2b |
| A10 | 876 | 8 | 28 | R | C | e | Kyoto | sunset | NSJ, v.29, ff.7a-7b |
| A11 | 876 | 10 | 16 | W | C | e-w | Kyoto | | NSJ, v.29, f.10a |
| A12 | 883 | 4 | 20 | W | C | n | Kyoto | near moon | NSJ, v.43, f.7b |
| A13 | 885 | 8 | 14 | B | C | en-ws | Kyoto | | NSJ, 48, f. 1b |

Abbreviations
Color: R = red, W = white, B = bluish
Description: V = vapor, C = cloud, L = light
Direction: e = east, w = west, s = south, n = north
Place: Asuka = Asuka Kyo, Kyoto = Heian Kyo

**Table 2: Japanese historical capital and locations of observatories**

| Since | Until | Location | | Geo. Latitude | Geo. Longitude |
|---|---|---|---|---|---|
| | 694 | Asuka | 飛鳥 | 34°29′ | 135°49′ |
| 694 | 710 | Fujiwara Kyo | 藤原京 | 34°30′ | 135°48′ |
| 710 | 740 | Heijo Kyo | 平城京 | 34°42′ | 135°48′ |
| 740 | 743 | Kuni Kyo | 恭仁京 | 34°46′ | 135°51′ |
| 743 | 744 | Shigaraki no Miya | 紫香楽宮 | 34°56′ | 136°05′ |
| 744 | 745 | Naniwa Kyo | 難波京 | 34°41′ | 135°31′ |
| 745 | 784 | Heijo Kyo | 平城京 | 34°41′ | 135°40′ |
| 784 | 794 | Nagaoka Kyo | 長岡京 | 34°56′ | 135°42′ |
| 794 | | Heian Kyo | 平安京 | 35°01′ | 135°45′ |